\documentstyle[aps,prl,amsmath,epsfig,twocolumn]{revtex}

\newcommand{\beps}{\boldsymbol{\epsilon}}
\newcommand{\bnh}{{\mathbf \hat n}}
\newcommand{\lpar}{lin\,\|\, lin}
\newcommand{\lperp}{lin\perp lin}
\newcommand{\hpar}{h\,\|\, h}
\newcommand{\hperp}{h\perp h}
\newcommand{\ket}[1]{|#1\rangle}
\newcommand{\bra}[1]{\langle#1|}
\newcommand{\mv}[1]{\langle#1\rangle}

\begin{document} 

\author{Thibaut Jonckheere$^{\dagger}$, 
Cord A. M\"{u}ller$^{\ast }$,
Robin Kaiser$^{\ast },$ 
Christian Miniatura$^{\ast }$
and Dominique Delande$^{\dagger}$ 
}
\address{$\dagger$ Laboratoire Kastler-Brossel, Universit\'e Pierre 
et Marie Curie\\
Tour 12, Etage 1, 4 Place Jussieu, F-75252 Paris Cedex 05, France\\
$\ast$ Institut Non Lin\'{e}aire de Nice, UMR 6618 du CNRS, 
1361 route des Lucioles,\\
F-06560\ Valbonne, France}
\title{Multiple scattering of light by atoms in the weak localization regime}
\date{\today}

\maketitle

\begin{abstract}
Coherent backscattering is a multiple scattering  
interference effect 
which enhances the diffuse 
reflection off a disordered sample in the backward direction.
Classically, the enhanced intensity is 
twice the average background under well chosen
experimental conditions.
We show how 
the quantum internal structure of atomic scatterers
leads to a significantly smaller enhancement. 
Theoretical results for double scattering in the weak localization
regime are presented  
which confirm recent experimental observations.
\end{abstract}



Standard cooling techniques permit to 
prepare optically thick media 
of cold atoms, like
Bose-Einstein condensates~\cite{bec}. 
In a sufficiently dense disordered atomic medium, light is expected to
be localized by interference, in analogy to Anderson localization of 
electrons in disordered
solid-state samples~\cite{anderson}.
For multiple scattering of light, atoms are usually
modeled
as dipole point scatterers, their
internal structure being ignored. 
In this Letter, we show
that even far from
strong localization, i.e. in a dilute medium,  
such an approximation 
is severely too optimistic 
and we 
demonstrate that the  
degeneracy of the atomic ground
state 
can significantly reduce 
interference effects.

When monochromatic light is elastically scattered off a disordered medium,
the interference between all partial waves 
produces strong angular fluctuations of the intensity distribution  
 known as a speckle pattern. 
In the weak localization regime (a dilute disordered medium 
such that $k\ell \gg 1$  
where $k=2\pi/\lambda$ is the light wave number and
$\ell$ the elastic mean free path)
-- the case we consider in 
the following  -- 
the phases associated with different scattering paths are essentially
uncorrelated. 
Averaging over the positions of the scatterers 
then washes out interferences and produces a smooth 
reflected intensity. 
There is, however, an exception: 
the ensemble average cannot 
wash out the
interference between a wave travelling along a scattering path and 
the wave travelling along the reverse path (where the same scatterers
are visited in reverse order). Indeed, 
in the backscattering direction, the optical lengths of the direct
and reverse paths are equal. 
Thus the two waves interfere constructively 
and enhance the 
average diffuse reflected intensity. 
This phenomenon is known as coherent backscattering (CBS), 
a hallmark of interference effects in disordered 
systems~\cite{sheng,wavediffusion}.   

The average intensity $I$ scattered at angle $\theta$ 
can be written as a sum of three terms, 
$I(\theta)=I_S(\theta)+I_L(\theta)+I_C(\theta)$
~\cite{vanderMark88}. 
Here, $I_S$ is the
single scattering contribution (for which the direct and reverse paths
coincide),  $I_L$ the raw contribution of multiple
scattering paths  (the so-called ladder terms) and $I_C$ the CBS
contribution (the so-called maximally crossed terms). 
$I_L$ and $I_S$ do not contain any interference term and thus
vary smoothly with $\theta$.
The contribution to $I_L$ of a pair
of direct and reverse paths is essentially 
$|T_{\mathrm dir}|^2+|T_{\mathrm rev}|^2$ 
where $T_{\mathrm dir}$ and $T_{\mathrm rev}$ are the corresponding 
scattering amplitudes. 
The contribution to $I_C$ is 
$2 |T_{\mathrm dir}| |T_{\mathrm rev}| \cos \phi,$ with 
$\phi = ({\mathbf k}+{\mathbf k'})
\cdot ({\mathbf r'}-{\mathbf r}),$
where ${\mathbf k},{\mathbf k'}$ are the incident and scattered
wave vectors and ${\mathbf r},{\mathbf r'}$ the positions of the first and
last scatterer along the path.
From these expressions, it follows that $I_C$ is always smaller or equal
than $I_L$. 
For a small scattering angle $\theta$, 
the phase difference 
$\phi$ 
is essentially 
$\theta k\ell$. 
Thus, $I_C(\theta)$ is peaked around the backscattering direction $\theta=0$
and rapidly decreases  to zero over an angular width $\Delta \theta
\sim 1/k\ell \ll 1$ \cite{Akkermans86}.

In usual experiments, the incident light is polarized either linearly or
circularly (with a given helicity $h$) and one studies the scattered
light with the same or orthogonal polarization in four
polarization channels: $\lpar$, $\lperp$, $\hpar$ (helicity is preserved)
and $\hperp$ (helicity is flipped). 
The ratio of the 
average intensity in the backward direction 
$I(\theta =0)$ to the 
background 
$I(1/k\ell \ll \theta \ll 1)=I_S+I_L$ defines the enhancement factor
in each polarization channel:
\begin{equation}
\alpha = 1+ \frac{I_C(0)}{I_S+I_L}.
\label{enhancementfactor}
\end{equation}
As $I_C\le I_L$, its largest possible value is 2, reached if and only
if 
$I_S=0$  
and $I_C=I_L$.  
For classical scatterers and exact backscattering
~\cite{mishchenko92}, the first condition $I_S=0$ is 
fulfilled in the $\lperp$ and $\hpar$ channels if the scatterers have
spherical symmetry. 
The second condition $I_C=I_L$ is fulfilled in the $\lpar$ and
$\hpar$ channels provided reciprocity holds. 
Reciprocity is a symmetry property   
valid whenever the fundamental
microscopic description of the system is time reversal invariant
\cite{onsager}. It assures that 
\begin{equation}
T_{\mathrm dir}({\mathbf k} \beps \rightarrow {\mathbf k}' \beps') =
T_{\mathrm rev}(-{\mathbf k}' {\beps'}^{*} \rightarrow -{\mathbf k} \beps^{*})
\label{reciprocity_cl}
\end{equation}
where $(\mathbf k,\beps)$ and $(\mathbf k',\beps')$ 
are the incident and scattered wave vectors and polarizations
(the star indicates complex conjugation). 
 In the backscattering direction 
$(\mathbf k'=-\mathbf k)$ and parallel polarization channels 
$(\beps'=\beps^*)$, the scattering amplitudes of any 
couple of direct and reverse paths are thus identical, 
leading to $I_C=I_L$~\cite{bart}. 
As a consequence, the maximum enhancement $\alpha=2$ is expected for 
spherical scatterers in
the $\hpar$ channel, a prediction confirmed by experiment
~\cite{wiersma95}.

Recently, CBS of light was observed with cold atoms 
with surprisingly low enhancement factors,
with the lowest value in the $\hpar$ channel~\cite{labeyrie,labeyrie2}.
The quantum internal structure of
the atoms can account for a major part of this astonishing observation. 
Two different reasons must be distinguished:     
$(i)$ all polarization channels now contain a single scattering
contribution, 
$(ii)$ the amplitudes interfering for CBS are in general not
reciprocal which leads to 
$I_C<I_L$.   
This Letter is devoted to elucidating these points and presents 
an analytical calculation of the CBS signal 
with atoms in the double scattering regime.
 
We consider a collection of atoms at rest exposed to monochromatic
light, quasi-resonant with an electric dipole transition between
some ground state with angular momentum $J$ and some excited
state with angular momentum $J'$.
For sufficiently weak light intensity, a perturbative
description is in order: 
an atom with initial state $\ket{Jm}=\ket{m}$
undergoes a transition into a final state 
$\ket{Jm'}=\ket{m'}$ while scattering  
an incoming photon (${\mathbf k},\beps$)
into an outgoing mode  (${\mathbf k}', \beps'$).  
We assume that no magnetic field is present, so that 
the atomic ground state is $(2J+1)$--fold degenerate. 
Energy conservation then implies that 
the scattering process is purely elastic  
$(\omega=\omega')$. 

The
single scattering transition amplitude $T_S$ 
is proportional to the 
matrix element  
$t_S(\beps,m;\beps',m')=
\bra{m'}({\beps'}^*\cdot {\mathbf d})(\beps\cdot{\mathbf d})\ket{m}$ 
of the resonant scattering tensor \cite{diagrams}
where $\mathbf d$ is the 
atomic dipole operator connecting the $J$ and $J'$ subspaces. 
This amplitude describes the absorption $(\beps\cdot{\mathbf d})$
of the incoming photon, and the emission
$({\beps'}^*\cdot{\mathbf d})$ of the final photon.
Now not only transitions to the same $m'=m$ 
substate are allowed (Rayleigh transitions), but also to 
$m'\ne m$ (degenerate Raman transitions).
It is then in general no longer possible to eliminate the single scattering
contribution by polarisation analysis. 
For example, a signal in the $\hpar$ 
channel in the backscattering direction
is associated by angular momentum conservation to a transition
$|m'-m|=2$ 
and can be suppressed only if $J=0,1/2$.

Of more fundamental interest is the second reason for the enhancement
reduction, $I_C<I_L$, and the role of reciprocity. 
As a general
rule in quantum mechanics, only transition amplitudes    
which connect the {\em same} initial
state to the {\em same} final state can interfere. 
Here the states of the complete system are
the photon modes {\em and}  
the  internal states $\ket{\{m\}}= \ket{m_1,m_2,\dots}$ of all atoms.
Again, CBS originates from the interference between 
two amplitudes 
$T_{\mathrm dir}$ and $T_{\mathrm rev}$ 
associated 
with the direct and reversed scattering
sequences of the same transition $\ket{{\mathbf
k}\beps,\{m\}}\rightarrow \ket{{\mathbf k}'\beps',\{m'\}}$.
But in this case, reciprocity -- although it perfectly holds -- 
fails to predict 
the enhancement factor.
Indeed, the reciprocity relation now 
reads~\cite{quantum_reciprocity}
\begin{align}
T_{\mathrm dir}({\mathbf k}\beps, & \{m\} \rightarrow  {\mathbf
k}'\beps',\{m'\}) = (-1) ^ {\sum_i(m_i' - m_i)} \nonumber  \\ 
& \times  T_{\mathrm rev}(-{\mathbf k}' {\beps'}^{*},-\{m'\}\rightarrow -{\mathbf k}
\beps^{*},-\{m\}).
\label{reciprocity_quant}
\end{align}
In the backscattering direction 
$(\mathbf k'=-\mathbf k)$ and parallel polarization channels  
$(\beps'=\beps^*)$ 
these two reciprocal amplitudes  will interfere and thus 
contribute to CBS if and only if 
$\{m'\}=-\{m\}$.  
So, among all pairs of interfering amplitudes, 
only a few are linked by reciprocity 
($m'=-m=0$ for Rayleigh transitions, $m'=-m=\pm 1,\pm 1/2$ for degenerate 
Raman transitions).   
Unless this condition is fulfilled, the reciprocal
amplitudes are associated with different initial
and final states of the system and cannot interfere.  
This is similar to the case of orthogonal polarizations where 
reciprocity indeed is still valid, but simply inapplicable. 
In the case of atomic scatterers with internal structure, 
this is true for all polarization channels, and stands out in sharp 
contrast to the classical case. 
We point out, however, that 
the condition $\{m'\}=-\{m\}$ is trivially true  
in the case of an 
$J=0\rightarrow J'=1$ transition since the ground state  
then has no internal structure and $m=m'=0$. 
Therefore, the enhancement factor for this transition will 
be the same as for classical point scatterers.

\begin{figure}
\centerline{\epsfig{figure=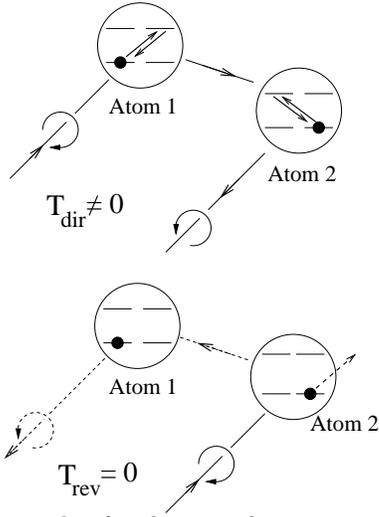,width = 5cm,angle=0}}
\caption{Example  
of a direct and reverse scattering path 
having different amplitudes: double Rayleigh scattering 
on a $J=1/2 \rightarrow J'=1/2$ 
transition in the helicity preserving channel with positive incident
helicity.
The arrows show atomic transitions corresponding to
absorption and emission of photons; 
the dashed lines show the process which has a vanishing amplitude.}
\label{figexample}
\end{figure}

Thus no fundamental reason is left for $I_C=I_L$ to hold. 
Let us illustrate why one expects now $I_C<I_L$ in general. 
Consider double 
scattering on two atoms without change of their internal states   
($m_1'= m_1$; $m_2'=m_2$), for $J=1/2 \rightarrow J'=1/2$ 
in the $\hpar$ channel with positive incident
helicity.  
The atoms are supposed to be  
initially prepared in the $|m_1=-1/2\rangle$ and $|m_2=+1/2\rangle$ 
substates 
(quantization axis parallel to the incoming light wavevector). 
In this configuration (see fig.~\ref{figexample}), atom $1$ can 
scatter the incident photon. 
The intermediate photon can be scattered 
by atom 2 to be detected in the backward direction with the required
helicity. 
Along the reverse path, the photon 
must be scattered first by atom 2. 
But
atom 2 cannot scatter the incident photon with positive helicity 
because it is in the maximum magnetic quantum number state $\ket{m_2=+1/2}.$ 
This simple example shows a situation where the  
reverse amplitude $T_{\mathrm rev}$ is strictly 
zero while the direct one $T_{\mathrm dir}$ is not. Consequently, this 
vanishing 
interference does not contribute at all 
to the CBS enhancement factor. 
More generally, a path and its reverse partner
will have non-zero but different amplitudes 
$T_{\mathrm dir}\ne T_{\mathrm rev}$, resulting in a loss of contrast
and an overall 
enhancement factor less than $2$.

We now sketch the general lines of the complete 
calculation. Consider double scattering 
(higher orders can be calculated similarly)
of a photon 
$({\mathbf k},\beps)$ into the mode $({\mathbf k}',\beps')$.  
The transition amplitudes 
can be calculated using standard diagrammatic
techniques
\cite{diagrams}.  
The dependence of the direct amplitude $T_{\mathrm dir}$
on the internal atomic structure 
factorizes into    
\begin{equation}
 t_{\mathrm dir}
= 
 \bra{m_1'm_2'}({\beps'}^*\cdot{\mathbf d}_2) 
({\mathbf d}_2\cdot {\mathbf \Delta}\cdot {\mathbf d}_1)
(\beps\cdot{\mathbf d}_1)
 |m_1,m_2 \rangle
\label{amplitude}
\end{equation} 
where ${\mathbf\Delta}_{ij}= \delta_{ij}-\bnh_i\bnh_j$ is the
projector onto the plane transverse to the unit vector $\bnh$ joining
the two atoms. 
The internal part of $T_{\mathrm rev}$ is obtained by 
exchanging ${\mathbf d}_1$ and 
${\mathbf d}_2$ in eq.~(\ref{amplitude}) 
(but {\em not} the magnetic quantum numbers):
\begin{equation}
 t_{\mathrm rev}
= 
 \bra{m_1'm_2'}({\beps'}^*\cdot{\mathbf d}_1) 
({\mathbf d}_1\cdot {\mathbf \Delta}\cdot {\mathbf d}_2)
(\beps\cdot{\mathbf d}_2)
 |m_1,m_2 \rangle
\end{equation}  

The average reflected intensity 
is proportional to 
$\mv{|T_S|^2}+\mv{|T_{\mathrm dir}+T_{\mathrm rev}|^2}$ 
where the brackets $\mv{\dots}$ indicate the
ensemble average 
over the spatial distribution of the atoms and the average over the 
distribution of internal states  $\{m\}$. 
We assume the atoms to be 
uniformly distributed 
in half-space, allowing comparison with results for classical 
point scatterers \cite{tiggelen90}. 
The initial distribution of internal states is supposed 
to be a complete statistical mixture, 
which is likely to be the case under usual experimental conditions.

Two different techniques have been used for the average over
internal states: firstly, we have calculated the direct and reverse
amplitudes for every possible initial and final state
in terms of Clebsch-Gordan coefficients and 
performed the sum over all possible states using a standard
calculation program (Mathematica).
Secondly, we derived analytical expressions~\cite{Mueller00} for the
single and double
scattering contributions using standard techniques of 
irreducible tensor operators \cite{oti}. 
The two approaches give the same result. 
Let us note that all results known for classical point
scatterers~\cite{tiggelen90} are recovered in the case of a $J=0\rightarrow
J'=1$ transition, for example an enhancement factor equal to 2 in the $\hpar$ 
channel.

\begin{figure}
\centerline{\epsfig{figure=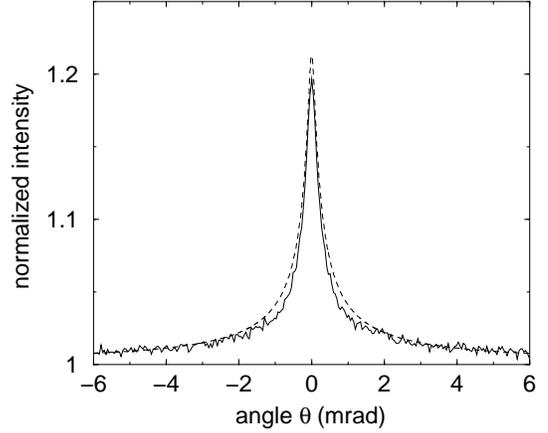,width = 7cm,angle=0}}
\caption{
Backscattered intensity normalized to the background  
as a function of the deviation $\theta$ from exact backscattering.
The solid line is the experimental 
observation~{\protect\cite{labeyrie,labeyrie2}} 
in the $\hperp$ channel
on Rb atoms ($J=3\rightarrow J'=4$ transition)
while the dashed line is the result of the calculation
taking into account single and double scattering and
the internal structure of the atom. The reduction of the enhancement
factor from 2 to 1.2 is due to both the loss of contrast 
in the interference between direct and reverse
scattering sequences because of the quantum internal structure of the
atomic scatterers, and the contribution of single scattering. }
\label{comparison}
\end{figure}

The normalized 
CBS angular intensity profile in the $\hperp$ channel
for a $J=3 \to J'=4$ transition is shown in fig.~\ref{comparison}
together with the experimental curve, see 
ref.~\cite{labeyrie,labeyrie2}. For the calculation, we used
the measured value $k\ell\approx 14,000$.    
The agreement is satisfactory, 
the shapes being similar, with  a calculated width 
$\Delta \theta=0.57\,$mrad, reasonably close to the
experimental value of $0.50\,$mrad.

In table~\ref{table}, we show the enhancement factors predicted by 
our calculation for the
transition $J=3 \rightarrow J'=4$ in the various channels together
with the experimentally observed values. We also show the value of the
enhancement factor when single scattering is not taken into account,
that is $1+I_C(0)/I_L$;
this is a direct measure of the effect of the internal structure
of the atom, as this quantity is equal to 2 in all channels for 
classical spherical scatterers (and for an $J=0 \to J'=1$ transition).
Clearly, in the $\hpar$ channel, the imbalance of interfering amplitudes 
appears as  
the key mechanism of the observed reduction of the enhancement factor.
In the other channels, the single scattering contribution also
plays an important role.

Few words of caution are necessary: we have here 
neglected higher scattering orders and assumed a semi-infinite
medium whereas the experimental cloud rather has a Gaussian distribution
of scatterers. A detailed quantitative comparison should take both
these effects into account and is under current study. 
It is thus not surprising that the 
calculated enhancement factors are only in fair agreement with the
experimental observation in some channels. However, the fact
that the calculation predicts semi-quantitatively correct results 
-- especially the surprising observation that the
maximum enhancement is obtained in the $\hperp$ channel and the
minimum one in the  $\hpar$ channel -- 
is a strong indication that we have caught the essential physical
mechanisms.

It should be noticed that scattering events which change the internal state
of an atom (degenerate Raman transitions) do contribute to the CBS
signal, as the direct and reverse paths lead to the
same final state (different from the initial state). In contrast,
the light scattered in a degenerate Raman transition does not interfere
with the incoming light simply because they are associated with different
final atomic states. There are even situations where the degenerate Raman
transitions dominate in the CBS signal: in the $\hpar$ channel
for a $J=3\to J'=4$ transition, more than 80\% of the double scattering
CBS signal originates from such transitions.

In conclusion, we have presented a calculation of the CBS
enhancement factor including single and double scattering
contributions for an atomic transition $J\to J'$.  
The main result is the
spectacular loss of contrast in the interference between direct and reverse
scattering sequences because of the quantum internal structure of the
scatterers.   
We think this could be of paramount
importance in the understanding of light propagation
in cold atomic media and in the search for strong localisation of light.

We thank Serge Reynaud, Jakub Zakrzewski and Bart van Tiggelen for numerous
fruitful discussions. Laboratoire Kastler Brossel de
l'Universit\'e Pierre
et Marie Curie et de l'Ecole Normale Sup\'erieure is
UMR 8552 du CNRS.

\begin{table}
\begin{tabular}{c|ccc} 
  &  Double scattering & Single and double  & Experimental  \\ 
  &  only & scattering  &  value  \\ 
     \hline
$\hperp \mbox{        }$&  1.71 & 1.21 & 1.20\\  
$\lperp \mbox{        }$&  1.45 & 1.20 & 1.15\\  
$\lpar \mbox{        } $&  1.60 & 1.20 & 1.12\\ 
$\hpar \mbox{       }  $&  1.22 & 1.17 & 1.06\\  
\end{tabular}
\caption{Enhancement factor  
of the average intensity scattered in the backward direction for a  
$J=3\to J'=4$ transition on atoms uniformly distributed in a 
semi-infinite medium,  
in the four polarization
channels. In the first column, single scattering is not included in
the calculation, so that
deviations from a factor 2 are entirely due to the the 
imbalance of interfering amplitudes along the direct and reverse paths.
For classical dipole point scatterers, this factor is 2 in the four channels. 
When single scattering is taken into account in the calculation
(second column), the values agree fairly well with the experiment.}
\label{table}
\end{table}

\end{document}